\documentstyle[aps,12pt]{revtex}
\newcommand{\lettersize}{\baselineskip=0.5cm}
\newcommand{\dir}{FIGS}
\input psfig
\input{psfig}
\newcommand{\fig}[3]
{
\begin{center}
     \noindent
     \unitlength=1mm
     \begin{picture}(#2,#3)
     \put(0,0){
       \psfig{figure=\dir/#1,width=#2mm,height=#3mm}
     }
     \end{picture}
   \noindent
\end{center}
}
\begin{document}
\baselineskip=12pt
\setcounter{page}{1}

\noindent
{\LARGE\bf
Monte Carlo simulations of copolymers at homopolymer interfaces:
Interfacial structure as a function of the copolymer density
}

\vspace{0.5cm}

\lettersize

\begin{center}
A. Werner, F. Schmid, M. M\"uller \\
{\em Institut f\"ur Physik, Universit\"at Mainz, D-55099 Mainz, Germany}
\end{center}


\begin{quote}
{\bf Abstract.}
By means of extensive Monte Carlo simulations of the bond fluctuation
model, we study the effect of adding AB diblock copolymers on the
properties of an interface between demixed homopolymer phases.
The parameters are chosen such that the homopolymers are strongly
segregated, and the whole range of copolymer concentrations in 
the two phase coexistence region is scanned. We compare the ``mushroom''
regime, in which copolymers are diluted and do not interact with 
each other, with the ``wet brush'' regime, where copolymers overlap
and stretch, but are still swollen by the homopolymers. A ``dry brush''
regime is never entered for our choice of chain lengths. 
``Intrinsic'' profiles are calculated using a block analysis method
introduced by us in earlier work. We discuss density profiles,
orientational profiles and contact number profiles.
In general, the features of the profiles are similar at all copolymer
concentrations, however, the profiles in the concentrated regime
are much broader than in the dilute regime.
The results compare well with self-consistent field calculations.
\end{quote}


\lettersize

\section{Introduction}

Polymers of different types are usually immiscible\cite{blends,blends2}. 
When blending different polymers together, one often obtains a 
mesoscopically structured material which contains numerous droplets of one
phase finely dispersed in the other. The structure
of the interfaces between the two phases, as well as the number and size of the
droplets thus crucially determine the properties of the composite material.
The droplet size, in turn, depends not only on the conditions of 
preparation and kinetic factors, but also
on the interfacial tension\cite{taylor,wu}.

Understanding the interfacial properties and, if possible, improving them
has therefore been a matter of longstanding interest\cite{interfaces,fracture}.
Interfacial properties can be tuned in a very efficient way by adding 
surfactants which adsorb at the interface. 
In the case of homopolymer blends, the most
natural surfactant molecules are copolymers, in which monomers of both
types are connected to each other by chemical links\cite{copo1,copo2,copo3}.

The effect of copolymers on homopolymer interfaces is twofold. 
First, they reduce the number of direct contacts between homopolymers, thereby 
reducing the interfacial tension\cite{leibler,sem3,agk,ramfs,ir,jw,lh,zh}. 
Second, they significantly enlarge the interfacial region, where polymers 
bridging the interface between the coexisting phases
can entangle, thereby increasing the fracture 
toughness\cite{lc,pjb}.  
Note that in pure homopolymer interfaces, the entanglements which
contribute to the mechanical stability
are those between polymers of different type, and are thus confined to
the narrow region where A and B polymers actually come into contact. 
If copolymers come into play, the relevant entanglements are those
between adsorbed copolymers and homopolymers, which are located in
a region with a width determined by the copolymer radius of gyration.

The energetic factors which favor the adsorption of copolymers at interfaces
are balanced against two opposing entropic factors: The loss of translational
entropy of copolymers and homopolymers, and at higher copolymer densities 
the elastic energy of stretching of the copolymer blocks. 
Following Leibler and Semenov\cite{leibler,sem3}, one can distinguish between 
four different regimes: At very low copolymer concentrations (mushroom regime),
the copolymers adsorb at the interface, arrange themselves such that each 
block sticks into its majority phase, and do not interact with each other. 
As the copolymer
concentration increases, the different blocks start to overlap and
the system enters the wet brush regime, where copolymers are forced to stretch, but remain 
swollen by the homopolymers. In the dry brush regime at even higher copolymer 
concentrations, the homopolymers are completely expelled from the 
central interfacial region. Finally, in the last regime, the interface
cannot accommodate any additional copolymers and saturates. 

On increasing the chemical potential of the copolymers, the excess 
concentration of copolymers at the interface increases and
one should thus pass through these different regimes. However, the
process is usually interrupted at some point due to the formation of
a new, copolymer-rich phase\cite{ir,hosch,mwm,philip}.
This phase can be either ordered\cite{mwm,philip} or disordered
but structured, {\em i.e.}, a microemulsion\cite{marcus1,bates,balsara}. 
This has been also observed by experiments of Balsara\cite{balsara} 
and co-workers and Bates {\em et al}\cite{bates}.

The structure of copolymers at interfaces has been studied intensely
in  experiments\cite{ramfs,lh,dai1,dai2,skht,bhp,gr,tprea,ma1,ma2,bud,smr,dnk}, 
and theoretically by various refined mean field 
theories\cite{noolandi,vilgis,shullk,shull,fith}. 
Computer simulations provide a useful way of obtaining additional 
structural information, and testing theoretical concepts\cite{simreview}.
Being computationally very demanding, only very few studies of homopolymer 
interfaces in the presence of copolymers exist so 
far\cite{marcus1,pan,mattice,psb,bgif,andreas1,jokim}. 

In our earlier work\cite{andreas1}, we have studied by Monte Carlo simulation the 
structure of symmetric diblock (AB) copolymers at a homopolymer (A/B) 
interface in the highly dilute mushroom regime within the bond fluctuation 
model (see section II). Our main results can be summarized as follows:
The copolymers in this non--overlapping regime assume the shape of
oriented dumbbells. The conformations of the individual blocks
resemble on average those of hardly perturbed homopolymer coils.
In particular, single blocks tend to show preferential alignment
parallel to the interface, even though the copolymers as a whole
orient perpendicular to the interface. Second, copolymers were
found to be significantly more compact than homopolymers at the 
center of the interface, {\em i.e.}, they have more intrachain contacts
and fewer interchain contacts. At distances of a few radii of gyration
from the interface, on the other hand, the opposite trend is found:
the number of interchain contacts increases at the expense of the self
contacts. In the present paper, we extend our previous work to the study 
of interfaces with higher copolymer content. 

The bulk phase behavior of ternary mixtures
of A,B homopolymers and symmetric (AB) diblock copolymers has been
investigated in detail by one of us\cite{marcus1}. Fig. 1 summarizes the 
resulting phase diagram (see section II for a description of the simulation 
model).  The addition of small amounts of copolymers has the effect of 
shifting the unmixing transition towards higher incompatibilities $\chi$.
At higher copolymer concentration, a tricritical point is encountered
and a miscibility gap opens up in which a copolymer rich phase,
either a microemulsion or at large $\chi N$ a lamellar phase, 
coexists with an A-rich or B-rich phase. The interfacial tension of the A/B 
interface was extracted by histogram reweighting methods, and the Fourier 
spectrum of capillary waves of an interface at moderate incompatibility
($\chi N = 9.2$) was analyzed in order to obtain some information
on the bending stiffness. It was found to be very small,
$\kappa/k_B T < 0.08$.

Here, we shall study the structure of the interface in microscopic detail,
for the case of strong segregation ($\chi N = 17.3$) and for copolymer
concentrations spanning the whole range from the dilute limit to the
point where phase separation sets in. Note that the coexisting
copolymer rich phase here is lamellar.
As in our previous work\cite{andreas1}, 
we shall focus on local properties, on profiles 
of local contacts and chain orientations, and local density profiles. We will 
compare the structure of the interface in the mushroom
regime and in the wet brush regime. The dry brush regime is
never reached, as we shall see: The homopolymers swell the copolymers
throughout the interface at all concentrations. 

In order to establish this last result in particular, it is necessary
to separate the local, intrinsic profiles from the capillary waves
of the interfacial position. This is done {\em via} a block
analysis procedure, which we have already applied successfully
to the study of homopolymer interfaces\cite{andreas2,marcus2,andreas3}. 
As far as possible, the results will be compared to the predictions of
a Helfand type self-consistent field 
theory\cite{noolandi,shull,helfand,friederike1}.

Our paper is organized as follows: In section II, we describe the
simulation model and the method and characterize
the system in somewhat more detail. The results are presented in
section III. First, we discuss the interfacial width and the interfacial
tension. Then, we study the effect of copolymer crowding on the
contact number profiles and the orientational properties. We summarize
and conclude in section IV.

\section{Model and technical details}


As in our earlier work, we use the bond fluctuation model, which is
a lattice model for polymer chains\cite{ck}. Three to five chemical repeat 
units are mapped onto one ``effective'' monomer, which occupies a cube of eight
lattice sites on a simple cubic lattice. Monomers are connected by
bonds of variable length $2$, $\sqrt{5}$, $\sqrt{6}$, 3, or $\sqrt{10}$. We 
work at the filling fraction 1/2, or monomer number density $\rho_b =1/16$, at 
which single chain configuration show almost ideal Gaussian chain statistics, 
{\em i.e.}, properties of a dense polymer melt are recovered\cite{pbhk}. 
The chain length is chosen as in Refs. [23,47], $N=32$. 
The chains then have the radius of gyration $R_g = b \sqrt{(N-1)/6} = 6.96$ 
with the statistical segment length $b=3.06$ in units of the lattice constants. 

To distinguish between monomers A and B, we introduce an
interaction potential acting between monomers with a distance of
less or equal $\sqrt{6}$ lattice constants:
$\epsilon_{AA} = \epsilon_{BB} = -\epsilon_{AB} = k_B T \epsilon$.
The parameter $\epsilon$ thus determines the relative repulsion
of unlike monomers. Here we use $\epsilon=0.1$, which is well in the regime 
where homopolymers demix ($\epsilon_c=0.014$\cite{hp1}). From $\epsilon$
one can estimate the Flory-Huggins parameter $\chi$ by \cite{marcus3}
\begin{equation}
\label{zeff}
\chi = 2 z_{\rm eff} \epsilon,
\end{equation}
where $z_{\rm eff}$ is the average number of interchain contacts of a
monomer. In our case, we measure $z_{\rm eff}=2.71$ in the demixed homopolymer 
bulk phase, which leads to $\chi = 0.54$.

We study ternary mixtures of A and B homopolymers and symmetric AB diblock
copolymers in the canonical and in the semi-grandcanonical ensemble. 
In the latter, the total number of polymers remains constant, but the polymers
can switch their identity between A or B homopolymer and copolymer.
The composition is then driven by two independent combinations of the chemical 
potentials\cite{marcus1}.
\begin{equation}
\label{chem}
\Delta \mu = \mu_A - \mu_B \qquad \mbox{and} \qquad
\delta \mu = \mu_C - \frac{1}{2}(\mu_A + \mu_B),
\end{equation}
where $\mu_A$ and $\mu_B$ are the chemical potential of the homopolymers , and $\mu_C$ denotes
the copolymer potential. Two phase coexistence between an A rich
and a B rich phase is found at  $\Delta \mu=0.$  The other variable
$\delta \mu$ determines the copolymer concentration.

In order to enforce a well-defined interface between
an A-rich and a B-rich phase, we place the system in a thick film of
thickness $D$ between asymmetric walls, one of which favors $A$ and
the other $B$. The wall potentials $\epsilon_w=0.1$ act on monomers
in the first two layers next to the wall, and are chosen strong 
enough that  each component wets its corresponding wall. Therefore, the
interface is located on average in the middle of the film. In the lateral
dimension $L$, periodic boundary conditions are applied. Specifically,
we use $L=128$ and $D=64$ or $D=128$, which ensures that the
film is thick enough that the interfacial properties are not affected
by the walls\cite{andreas2}. The system is equilibrated in 
the semi-grandcanonical ensemble. The Monte Carlo moves include
single monomer hopping moves, slithering snake moves\cite{kb},
identity switches of polymers, and polymer exchanges. 
After $ 2 \times 10^6$ Monte Carlo steps (MCS) 
which is well after the correct bulk concentrations of 
homopolymers and copolymers have been reached in the wings of the profiles, 
we turn off the identity switches and start to collect data.
Here four Monte Carlo steps correspond to one local hopping attempt per
monomer, three slithering snakes trials per chain and 0.1 
canonical particle exchange moves.
Our results thus refer to the canonical ensemble. The samples include
every 5000st configuration in runs of total length $2 \times 10^6$ MCS,
{\em i.e.}, 400 configurations.

``Intrinsic profiles'' profiles are obtained by a block analysis
which we have already applied successfully to homopolymer 
interfaces\cite{andreas2,marcus2,andreas3}.
The system is divided into blocks of size $B \times B \times D$ and
the interface position $h(x,y)$ is determined in each block.
This allows to study the capillary waves of the interface position
$h(x,y)$ (see section 3.1) and to average over local profiles
relative to the local interface position. We shall use the
block size $B=8$ which was found to be a suitable choice at $N=32$ and 
$\epsilon=0.1$ in our earlier studies (see Ref. [51] 
for a detailed discusstion of this ``intrinsic coarse graining length'').

Our results are compared to self-consistent field 
calculations\cite{helfand,friederike2}. 
Polymers are described as continuous space curves $\vec{r}(s)$, $s \in [0:N]$
with statistical weight
\begin{equation}
\label{gauss}
{\cal P}\{ \vec{r}(\cdot)\} = {\cal N}
\exp \big[ -\frac{3}{2 b^2} \int_0^N ds \Big|\frac{d \vec{r}}{ds}\Big|^2 \big]
\end{equation}
in the inhomogeneous external field 
$\omega_{\alpha}(\vec{r}) 
= \delta \beta {\cal F}/\delta \rho_{\alpha}(\vec{r})$
which is created by a monomer interaction potential
of Helfand type\cite{helfand}
\begin{equation}
\label{helfand}
\beta {\cal F} = \frac{1}{\rho_b}
\int d\vec{r} \: \big\{
\chi \rho_A(\vec{r}) \rho_B(\vec{r}) + \frac{\zeta}{2} (\rho_A+\rho_B-\rho_b)^2
\big\}.
\end{equation}
The parameter $\zeta = 1/(\rho_b k_B T \kappa)$ is an inverse compressibility.
The global compressibility of an athermal melt ($\epsilon = 0$) has
been determined earlier\cite{marcus4}, leading to $\zeta = 4.1$. However, 
we have found in our simulations of homopolymer interfaces that
the local density profiles are driven by a local compressibility which is
much higher, leading to $\zeta = 1.9$\cite{andreas3}. Note that most quantities
depend only very slightly on  $\zeta$. We shall use $\zeta=1.9$ in
the following. The single chain partition function for homopolymers ($j=A,B$)
and copolymers ($j=C$) is then given by the path integral 
\begin{equation}
\label{qq}
{\cal Q}_j = 
\int {\cal D}\{\vec{r}(\cdot)\}
{\cal P}\{ \vec{r}(\cdot)\} 
\textstyle
\exp \big[- \int_0^N \sum_{\alpha=A,B} ds \:\omega_{\alpha}(\vec{r}(s)) 
\gamma_{\alpha,j} (s) \big],
\end{equation}
where $\gamma_{\alpha,j}(s)=1$ if site $s$ on a polymer is occupied by
an $\alpha$ monomer, and 0 otherwise. (In copolymers, for example,
$\gamma_{A,C}(s) = 1$ for $s \in [0:N/2]$, and $\gamma_{B,C}(s) = 1$ for 
$s \in [N/2:N]$.) The relative density of copolymer monomers is adjusted
with the parameter $\delta \mu$
\begin{equation}
\label{rho}
\rho_{\alpha} = - 
\sum_{j=A,B} \frac{\delta {\cal Q}_j} {\delta \omega_{\alpha}(\vec{r})} 
- e^{\delta \mu} 
\frac{\delta {\cal Q}_C} {\delta \omega_{\alpha}(\vec{r})}.
\end{equation}
After having solved the set of equations (\ref{helfand}), (\ref{qq}) and 
(\ref{rho}) self-consistently, the interfacial tension is evaluated
according to
\begin{equation}
\sigma = -\frac{1}{\rho_b} \int_{-\infty}^{\infty} \!\! d z \:
\{ \; \chi [\rho_A \rho_B - \rho_b^2 (1-m_b^2)/4] + 
\frac{\zeta}{2} 
[(\rho_A + \rho_B)^2 - \rho_b{}^2] \; \},
\end{equation}
where $m_b$ is the bulk value of the order parameter of demixing, 
$m = (\rho_A-\rho_B)/(\rho_A+\rho_B)$. In the following lengths
shall often be given in units of $w_{\rm SSL} = b/\sqrt{6 \chi}$ which
is the width predicted by the self-consistent field theory in
the limit of an incompressible binary mixture of infinitely long
homopolymers\cite{helfand}.

Fig. 2 shows the bulk copolymer concentration $\rho_{b,C}$ in the two
phase region as a function of the copolymer chemical potential 
$\delta \mu$ \cite{marcus1}. 
In the limit of small copolymer concentrations, the latter should be 
well approximated by
\begin{equation}
\label{dmu}
\delta \mu = \ln (\rho_{b,C}/\rho_b) + \chi N/2,
\end{equation}
where the first term accounts for the translational entropy of the 
copolymer, and the loss of translational entropy for the homopolymers,
and the second term describes for the enthalpic repulsion between 
homopolymers and copolymers. Fig. 2a,  however, illustrates that
eqn. (\ref{dmu}) does not describe the simulation data very well. 
On the other hand, excellent agreement can by reached if one assumes
that the value of $\chi$ for copolymer/homopolymer interactions differs from 
that for homopolymer/homopolymer interactions. The best fit value for 
$\epsilon=0.1$ is $\chi_{cop}=0.48$ or alternatively (using eqn. (\ref{zeff}))
$z_{\rm eff,cop}=2.42$. We have measured the copolymer contact number
in the bulk phase and indeed found $z_{\rm eff,cop}=2.3 \pm 0.1$ (cf. Fig. 10b).
The difference to the average homopolymer contact number
($z_{\rm eff}=2.71$) is probably caused by a difference of conformations
of minority blocks and majority blocks in the homopolymer bulk phase.
Earlier investigations have shown that the radius of gyration of 
homopolymer coils in their minority phase is reduced\cite{sb0,cifra,marcus5b,marcus5}. 
Likewise, the minority blocks in the copolymers shrink to some extent which 
implies that they have fewer interchain contacts at the expense of loosing some conformational entropy\cite{marcus5b,marcus5}.
Similar effects have recently been reported 
experimentally\cite{balsara,ghfneu}.

One could now argue that consequently, different values of $\chi$ should be
used for copolymer/homopolymer interactions and homopolymer/homopolymer
interactions in the self-consistent field formalism. However, $z_{\rm eff}$
varies strongly in the interfacial region, reaching values as high as
$z_{\rm eff}=3$ in the wings of the profiles (cf. Fig. 10b). This strong
position dependence cannot be derived self-consistently within our
simple formulation of the theory. Hence we abide by the most
idealizing approximations, taking $z_{\rm eff}=2.71$ independent of
monomer type and position.

The area density of excess copolymers $\nu$ at the interface as a function of
$\delta \mu$ is shown in Fig. 3
\begin{equation}
\nu = \frac{ n_C - \rho_{b,c} L^2 D/N }{L^2} .
\end{equation}
The results compare quite well with the self-consistent field prediction.
The extent of overlap between copolymers is reflected by the
quantity $\delta A = n_C \pi R_g^2/L^2$. In the most dilute system
$\delta \mu = 0$, we obtain $\delta A = 0.65$, {\em i.e.}, the copolymers
do not overlap. In the system with the highest copolymer concentration ($\delta \mu=3$), the copolymers
overlap strongly, $\delta A = 3.8$.

\section{Results}

\subsection{Interfacial width and interfacial tension}

Fig. 4 shows various monomer density profiles in the dilute mushroom
regime ($\delta \mu = 0 $) and at the highest copolymer concentration
before phase separation sets in ($\delta \mu = 3$). For comparison,
the A and B profiles of pure homopolymer phases (from Ref. [51])
are also included. In the dilute case, the total profiles of A or B
monomers are not affected by the presence of the copolymers\cite{andreas2}. 
In the dense case, they broaden significantly, growing from 
$w=1.1 w_{\rm SSL}$ to $w=1.5 w_{\rm SSL}$ in units of 
$w_{\rm SSL}=b/\sqrt{6 \chi}$. Even wider is the total interfacial region. 
The width of the segregation profiles of copolymers is roughly 
$6 w_{\rm SSL}$ at chain lengths $N=32$, {\em i.e.}, 
1.5 times the radius of gyration. As discussed in the introduction, this
is the width which actually determines the number of 
interface-strengthening entanglements.
Even at the highest densities, the homopolymers are never fully expelled 
from the interfacial region. Thus the dry brush limit is never reached, 
the formation of the lamellar phase sets in in the regime where the copolymers
still aggregate into a wet brush.

The individual segment profiles of end monomers of type A, $\rho_{e,A}$,
of A monomers in the middle of the A block $\rho_{1/4,A}$ and of
A monomers in the chain middle $\rho_{1/2,A}$ are compared to the
average over all bonds in Figure 5. Each distribution is normalized to unit area. As expected, the copolymer
joint profiles $\rho_{1/2,A}$ is centered around the middle of
the interface, and the further one moves towards the end of the chains,
the deeper the corresponding segment profile reaches into the A phase.
The profile over the monomers in the block middle reproduces quite
closely the average over all bonds. The agreement with the self consistent
field predictions for the shape of the individual segment distributions is very good.

Now the interfacial width can be defined in different ways:
\begin{itemize}
\item[a)] From a fit of the total order parameter profile 
$m(z) = \big[\rho_A(z)-\rho_B(z)\big]/\rho(z)$ to a tanh function:
$m(z) = m_b \tanh(z/w)$.
\item[b)] From a fit of the copolymer joint profile to a 
Gaussian distribution
$\rho_{1/2,i}  \propto \exp(- \pi z^2/w_c^2)$,
with $i=$ A or B \cite{FN1}.
\item[c)] 
From the excess internal energy $e_s$ at the interface {\em via}
$w_e = 2 e_s/(\rho_b \chi m_b{}^2)$\cite{marcus1}. 
\end{itemize}
Fig. 6 shows the results obtained with these different methods.
The width of the copolymer joint profile follows the intrinsic width
$w$ of the AB profiles closely, but is larger by a constant factor of
approximately 1.85. The estimate $w_e$ is reasonably close to $w$.
Also shown is the apparent width $w(B=L)$ obtained at block size
$B=L=128$. It grows much faster than $w$ with increasing copolymer density
which is a direct consequence of the fact that the interfacial tension
decreases rapidly and capillary waves become more and more important for the width
of the apparent profile.

This is illustrated in more detail in Figs. 7 and 8. Fig. 7 shows for
different copolymer concentrations the interfacial width $w$ in units
of $w_{\rm SSL}$, as obtained from a block analysis as a function of block
size $B$. As one would expect from Fig. 6, the width grows faster with
$B$ for large copolymer concentration. The effect can be used to
extract the interfacial tension from the simulation data\cite{andreas2}.
In order to do so, however, one needs to discuss the effect of the
bending stiffness on the capillary wave fluctuations which also
increases with increasing copolymer content. Let $h(x,y)$ denote the
local position of the interface. The capillary wave Hamiltonian
then reads\cite{capillary}
\begin{equation}
\beta {\cal F} =
\int \! dx \; dy \; \Big\{ \frac{\sigma}{2}
| \nabla h(x,y) |^2 + \frac{\kappa}{2} | \Delta h(x,y)|^2 \Big\},
\end{equation}
where $\sigma$ is the interfacial tension, and $\kappa$ the bending
rigidity. From this expression one derives the thermal distribution of
interface positions
\begin{equation}
\label{pp}
P(z) = \langle \delta(z-h(x,y)) \rangle
= \frac{1}{2 \pi s^2} \exp(-z^2/2 s^2)
\end{equation}
\begin{equation}
\mbox{with} \qquad
\label{ss}
s^2 = \frac{1}{4 \pi \sigma} 
\ln\big( \frac{\kappa + q_{min}^{-2} \sigma}{\kappa +q_{max}^{-2}\sigma}
\big),
\end{equation}
where the lower cutoff for blocks of size $B$ is given by $q_{min}=2 \pi/B$,
and the upper cutoff refers to the internal coarse graining length $B_0$,
namely $q_{max} = 2 \pi /B_0$. The description is of course only valid
for block sizes $B > B_0$. In that case, one can approximate apparent
profiles $\rho(z)$ by the convolution of intrinsic profiles $\rho_0(z)$
with the height distribution $P(z)$ (\ref{pp}). We have used this
expression to analyze the curves in Fig. 7. It turns out that the effect
of the bending rigidity on the interfacial width is quite small,
$\kappa < 0.1$ for all curves, and that it may safely be neglected.
Eqn. (\ref{ss}) with $\kappa =0$ was then used to extract the
interfacial tension. The results are shown in Fig. 8. The interfacial
tension decreases to 25 \% of its original value before phase separation
sets in. The values are in good agreement with independent values
obtained from bulk simulations by means of histogram reweighting
techniques in Ref. [23].

The compatibilizing copolymers also affect the total
density profile. At pure homopolymer interfaces, it exhibits a
pronounced dip at the center of the interface, since reducing the
total density is one (relatively expensive) way to reduce the number
of unfavorable contacts there. When copolymers are added, the dip
of the total density becomes smaller. Homopolymers, on the other
hand, are more and more expelled from the interfacial region (Fig. 9).

\subsection{Contact numbers and orientational properties}

When looking at the contact number profiles, we recover the
trends reported in our earlier study of the mushroom regime\cite{andreas1}.
According to the usual mean field assumption, the number
of self contacts should scale like $N_{\rm self} \propto \rho$ and
the number of interchain contacts like $N_{\rm inter} \propto \rho^2$,
{\em i.e.}, the effective coordination number is approximated by
$z_{\rm eff} \propto \rho$ independent of the monomer and chain type. 
Fig. 10 shows the deviations from this ideal behavior.

It turns out that homopolymer and copolymer contact numbers differ
markedly from each other. The strongest position dependence is found
for copolymers. The relative number of interchain contacts is reduced at 
the center of the interface and enhanced at distances of up to 
one end-to-end radius from the interface. This is because the
copolymers at this distance are still in contact with the
interface, and get pulled towards it by the end which sticks
into the opposite phase. As a consequence, they stretch and offer more
contact area. The number of self contacts decreases accordingly (Fig. 2).
The profiles are thus governed by two different length scales --
the interfacial width and the radius of gyration of the copolymers.
At higher copolymer concentration, these effects are still present, 
but mellowed. The copolymers are less compact at the center of the interface 
than dilute copolymers, and less stretched in the wings. Qualitatively,
the contact number profiles of copolymers are similar in the mushroom 
regime and in the brush regime.

In the case of homopolymers, the contact number profiles change
qualitatively as the copolymer concentration increases. In the
mushroom regime, they reflect only one length scale, the interfacial
width. The effective coordination number decreases in the interfacial
region (cf. Refs. [47,68]), and the number of self contacts
increases. Note the fine structure at the center of the interface.
It is also found in pure homopolymer interfaces and has been
discussed in Ref. [51]. 
At higher copolymer concentrations, the contact number profiles
show the signature of a second length scale which is presumably
the copolymer radius of gyration. Copolymer and homopolymer profiles
now influence each other. The total net excess of $N_{\rm self}/\rho$ 
and the negative excess of $N_{\rm inter}/\rho^2$, however,
hardly depend on the copolymer concentration.

The total number of AB contacts per monomer depends
only very little on the copolymer concentration (Fig. 11).
The advantage of adding copolymers is that
they reduce the number of AB contacts between {\em homo}polymers,
as shown in Fig. 12.

Finally, we discuss the effect of copolymer overlapping on the orientational
properties of the copolymers and homopolymers. The squared end-to-end vector
components parallel ($x,y$) and perpendicular ($z$) to the interface 
are shown for homopolymers and copolymers in Fig. 13 and compared
to self consistent field predictions. Fig. 14 shows for comparison
the same quantities for single copolymer blocks, $R_{b,ee,i}{}^2$.
It is found that homopolymer coils close to the interface 
align to the interface, but not to the same
extent in the copolymer rich regime than in the dilute regime.
Likewise, the orientational tendencies of single copolymer blocks 
at the interface and away from the interface become weaker. It is
interesting to note that copolymer blocks centered right at the interface
still have a slightly parallel orientation.
However, the number of blocks centered at some distance from the 
interface, where the average orientation is perpendicular, is now much larger. 
The average value of $\langle R_{b,ee,i}{}^2 \rangle$
at the interface with the highest possible copolymer density, $\delta \mu=3$,
is $\langle R_{b,ee,z}{}^2 \rangle = 20.0 $ in the direction perpendicular
to the interface, as opposed to $\langle R_{b,ee,x}{}^2 \rangle = 15.4 $ 
in the direction parallel to the interface.
Hence the net orientation is perpendicular. Note that the copolymer blocks 
are on average slightly stretched.
For comparison, we give the corresponding values in the dilute case 
($\delta \mu = 0$): 
$\langle R_{b,ee,z}{}^2 \rangle = 17$ and $\langle R_{b,ee,x}{}^2 \rangle=16$,
{\em i.e.}, the copolymer blocks are hardly stretched or oriented at all.

When looking at the end-to-end vector of whole copolymers, one the additional 
effect comes into play, that the different blocks arrange 
themselves such that they stick into their majority phase.
Thus molecules as a whole are orientated perpendicularly at all copolymer
concentrations. However, in the dilute case, the orientation almost disappears 
for copolymers centered at the middle of the interface.
In the dense case, it is close to constant over a region of several 
radii of gyration's width (Fig. 13).


\section{Conclusions}

In summary, we have presented results of a large scale Monte Carlo simulation
aimed towards an understanding of the modification of local structure of an 
interface between immiscible homopolymers due to the presence of copolymers.
We have restricted ourselves to symmetric $AB$ copolymers, comprising the same
type of monomers as the corresponding homopolymers. Moreover all polymer species have 
identical degree of polymerization. Extending previous simulations\cite{andreas1} on this ternary 
melt we complement our knowledge of the phase diagram\cite{marcus1} and thermodynamical properties 
with a detailed analysis of the local structure of the interface at high segregation. 
The simulation results cover a broad range of concentrations ranging 
from the very dilute regime (mushroom regime) to the wet brush
regime. In the latter the copolymers crowd at the interface and reduce the concentration
of homopolymers at the center. However, the copolymers stretch only slightly and before
we reach the dry brush regime, where the copolymers stretch significantly and the homopolymer
density vanishes at the center of the interface, we encounter a first order transition to a lamellar,
copolymer rich phase. This is in agreement with SCF calculations\cite{philip} and experimental studies\cite{bates,balsara}. To reach the 
dry brush regime we presumably have to increase the segregation still further and increase the chain length 
of the homopolymers with respect to the copolymer.

Upon increasing the copolymer concentration in the bulk phases, the areal 
density of amphiphilic molecules adsorbed at the interface increases. 
The presence of the copolymers reduces the interfacial tension and increases 
the width of the composition profile. The broadening of the apparent 
interfacial profile in the Monte Carlo simulation is due to an enhancement of 
capillary waves caused by the reduction of the interfacial tension, and to the 
increase of the width of the ``intrinsic'' interfacial profile. 
The broadening of the apparent profiles has been investigated in detail by 
analyzing the interfacial profiles on different lateral coarse graining length 
scales $B$. This analysis yields an estimate of the interfacial tension which 
agrees well with independently measured values and shows that the bending 
rigidity of the interface remains quite small for the chain length investigated 
even at rather high areal density of copolymers. Choosing the lateral coarse 
graining length $B$ such that the width of the composition profile of the pure 
homopolymer interface agrees with its value in the SCF calculations, we have
extracted ``intrinsic'' properties of the interfacial profiles. 
This procedure allows us to distinguish between the effect of capillary waves 
and modification of the ``intrinsic'' interfacial properties, and we compare 
our simulational results to the SCF calculations without adjustable parameter.

The calculations in the framework of the Gaussian chain model capture the 
dependence of the interfacial properties on the copolymer concentration 
qualitatively and for some quantities --{\em e.g.} the shape of the 
distribution of individual segment as a function of the distance from the 
interface -- nearly quantitative agreement is achieved. However,
other quantities -- like the adsorbed amount at the interface as a function 
of the copolymer concentration in the bulk -- deviate quantitatively from the 
SCF predictions. At the rather low temperature studied, the system is well 
segregated and local composition fluctuations, which are neglected in the 
SCF treatment, cannot account for the deviations. Our simulation
results indicate that the conformation of the molecules depends on their
local environment -- {\em e.g.}, in their minority phase, they shrink in order 
to reduce the number of unfavorable intermolecular contacts and exchange them 
for energetically favorable intermolecular contacts. This leads to an
intricate interplay between molecular conformations and the distance from the
interface. The orientation of the copolymer's end-to-end vector is parallel 
to the interface, while the individual block orient perpendicular like the 
homopolymer coils. Moreover, the dependence of the molecular interaction 
energy is investigated in detail as a function of the distance from the 
center of the interface. The SCF calculations capture many but not all 
conformational and orientational effects at the interface.

The chain length dependence of our results has not been assessed. It is an 
open question to which extent these conformational changes will persist 
in the long chain length limit. We speculate that the orientation of the 
chains as a whole and the dumb-bell shape or polarization of the copolymers
at the interface will persist, because the entropy costs per molecule is of 
the order $k_BT$, hence chain length independent. This is also in agreement 
with recent experimental findings\cite{ghfneu,balsara}.
The reduction of the effective Flory-Huggins parameter of the copolymers in 
the bulk phases, and the concomitant deviations from the absorption isotherm, 
however, is caused by a shrinking of the minority block. The corresponding 
entropy loss per molecule increases with chain length (at $\chi N$
held constant) and simulations as well as analytical 
calculations\cite{marcus5b,marcus5} of binary homopolymer blends indicate that
this effect will decrease slowly with growing chain length.

\section*{Acknowledgments}

We wish to thank Philipp Janert, Michael Schick, and Thomas Veitshans
for useful discussions. Generous computing time was provided by the
ZDV, Mainz, the RHR, Kaiserslautern, and the Cray T3E at HLR, Stuttgart, and
at the HLRZ, J\"ulich. Partial financial support by the Deutsche 
Forschungsgemeinschaft under grant Bi 314/3-4 and Bi 317, by the 
Materialwissenschaftliches Forschungszentrum 
Mainz (MWFZ), and by the Graduiertenkolleg on supramolecular systems in Mainz
is gratefully acknowledged.


\begin{figure}
\fig{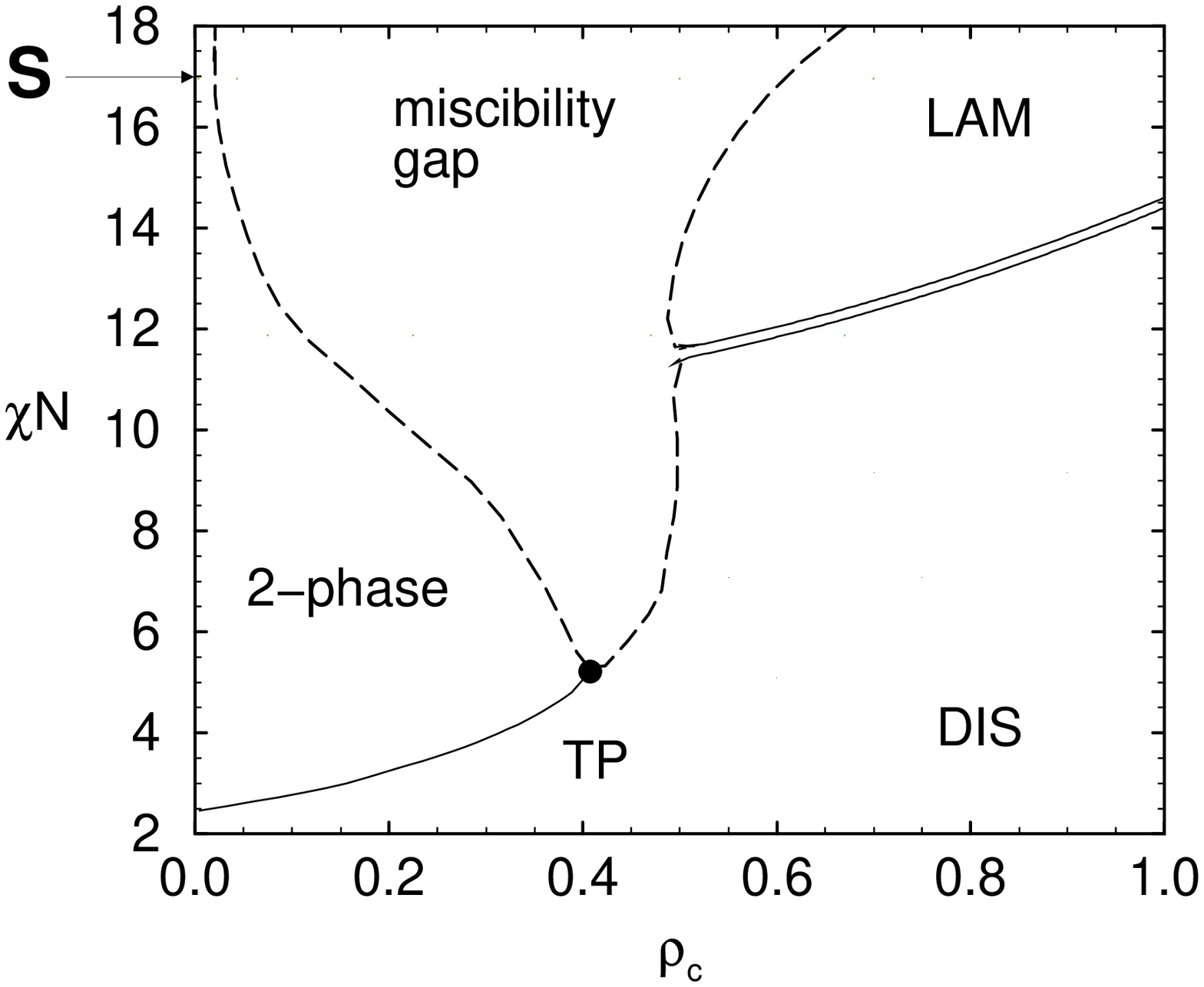}{100}{80}
\caption{ \label{fig1}}
Phase diagram of a symmetric ternary mixture of A and B homopolymers and
AB diblock copolymers of the same chain length $N=32$ in the bond fluctuation 
model (from Ref. [23]).
The exact location of the transition between the lamellar phase LAM
and the disordered phase DIS has not been determined; the double solid lines
are schematic. 
The simulations presented here were performed at $\chi N=17$
in the two-phase region (arrow).
\end{figure}


\begin{figure}
\fig{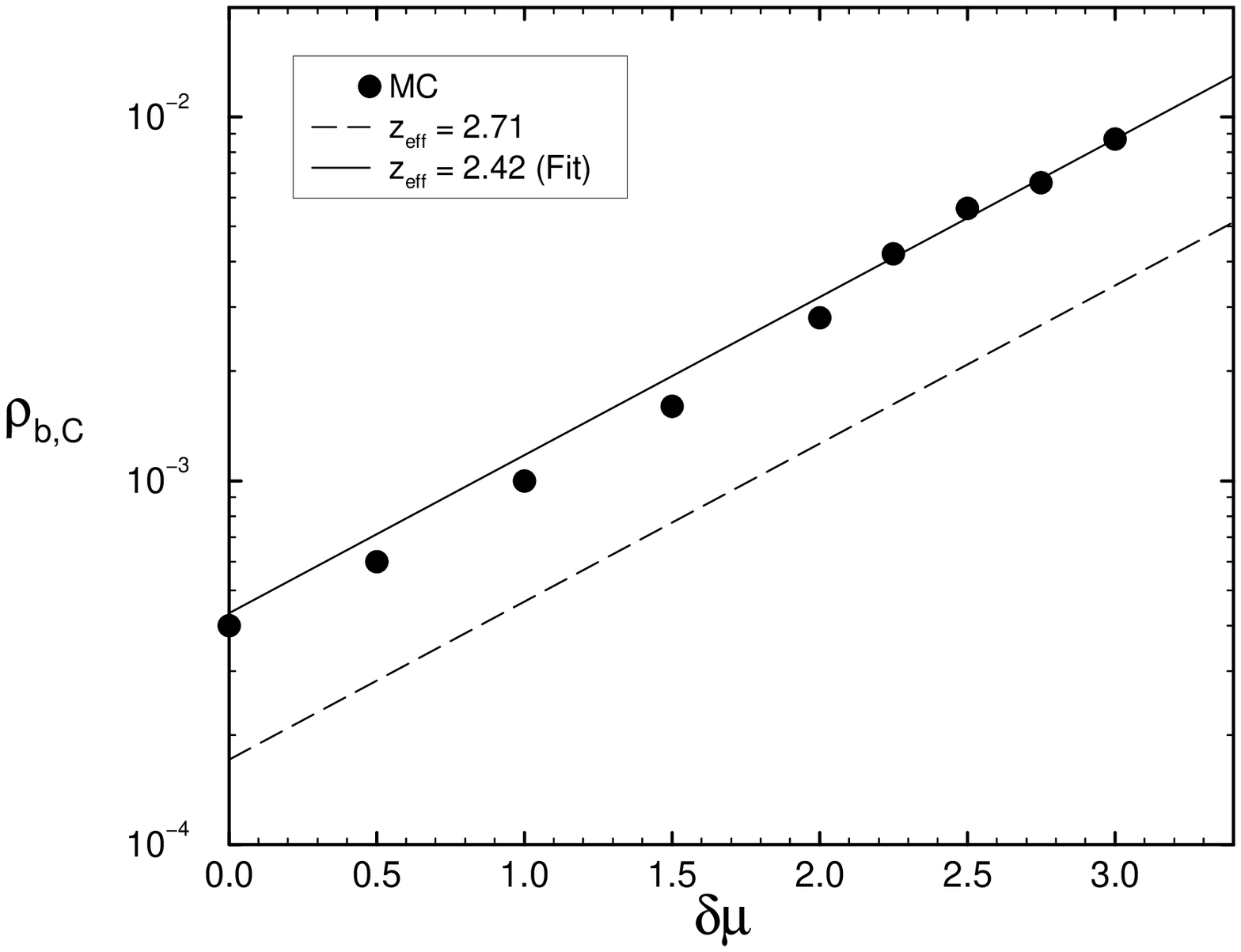}{100}{80}
\caption{ \label{fig2}}
Solubility of copolymers $\rho_{b,C}$ in the homopolymer phases
as a function of $\delta \mu$ at $\epsilon=0.1$, $N=32$ and $\Delta \mu=0$.
Symbols denote Monte Carlo results from Ref. [23].
Lines show the predictions of eqn. (\ref{dmu}) using $z_{\rm eff}=2.71$
(dashed) and $z_{\rm eff}=2.42$ (solid).
\end{figure}


\begin{figure}
\fig{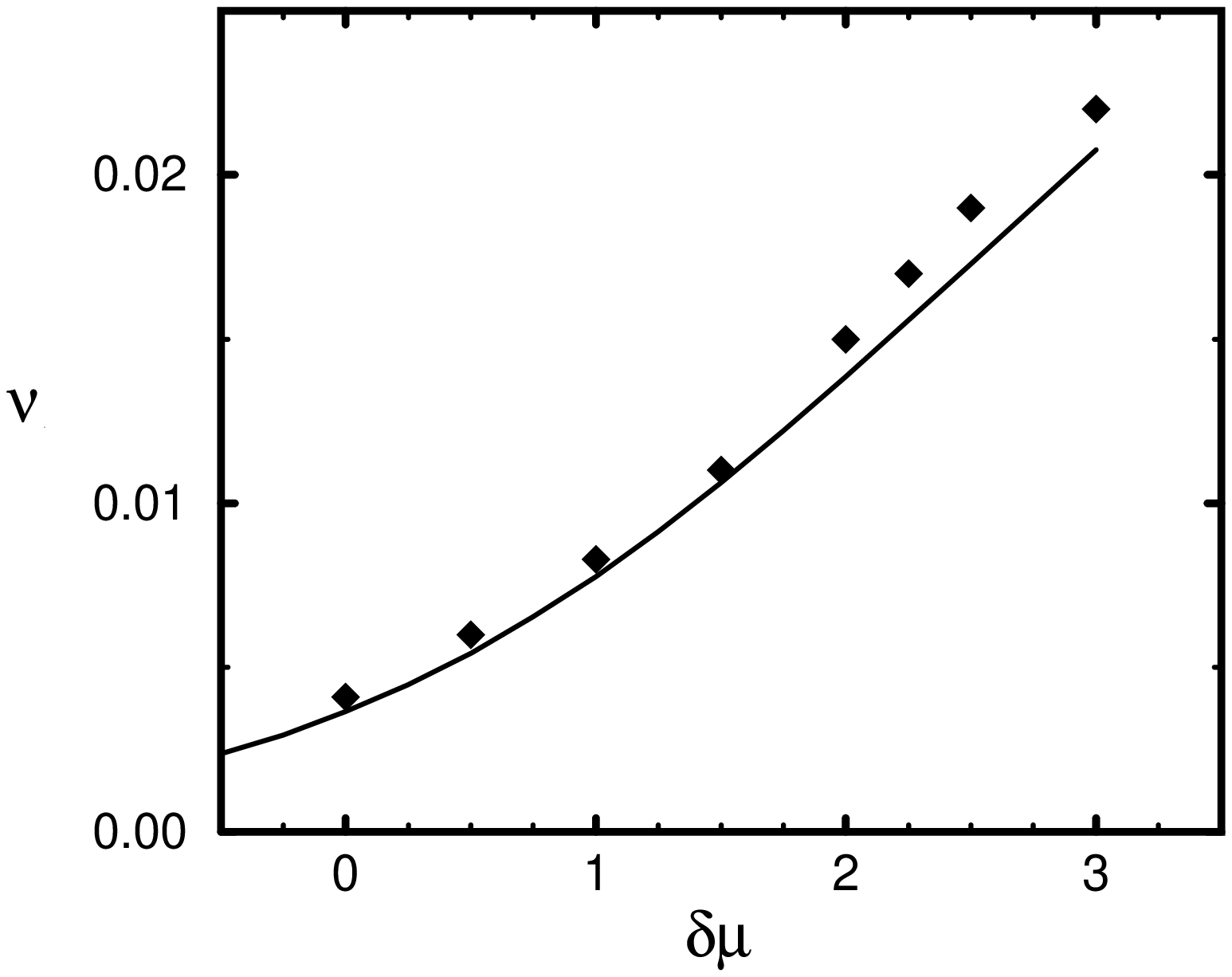}{100}{80}
\caption{ \label{fig3}}
Copolymer excess $\nu$ at the interface as a function of $\delta \mu$.
Solid line shows the prediction of the self-consistent field theory;
filled symbols denote the Monte Carlo results 
from Ref. [23].
\end{figure}

\clearpage

\noindent
(a)\fig{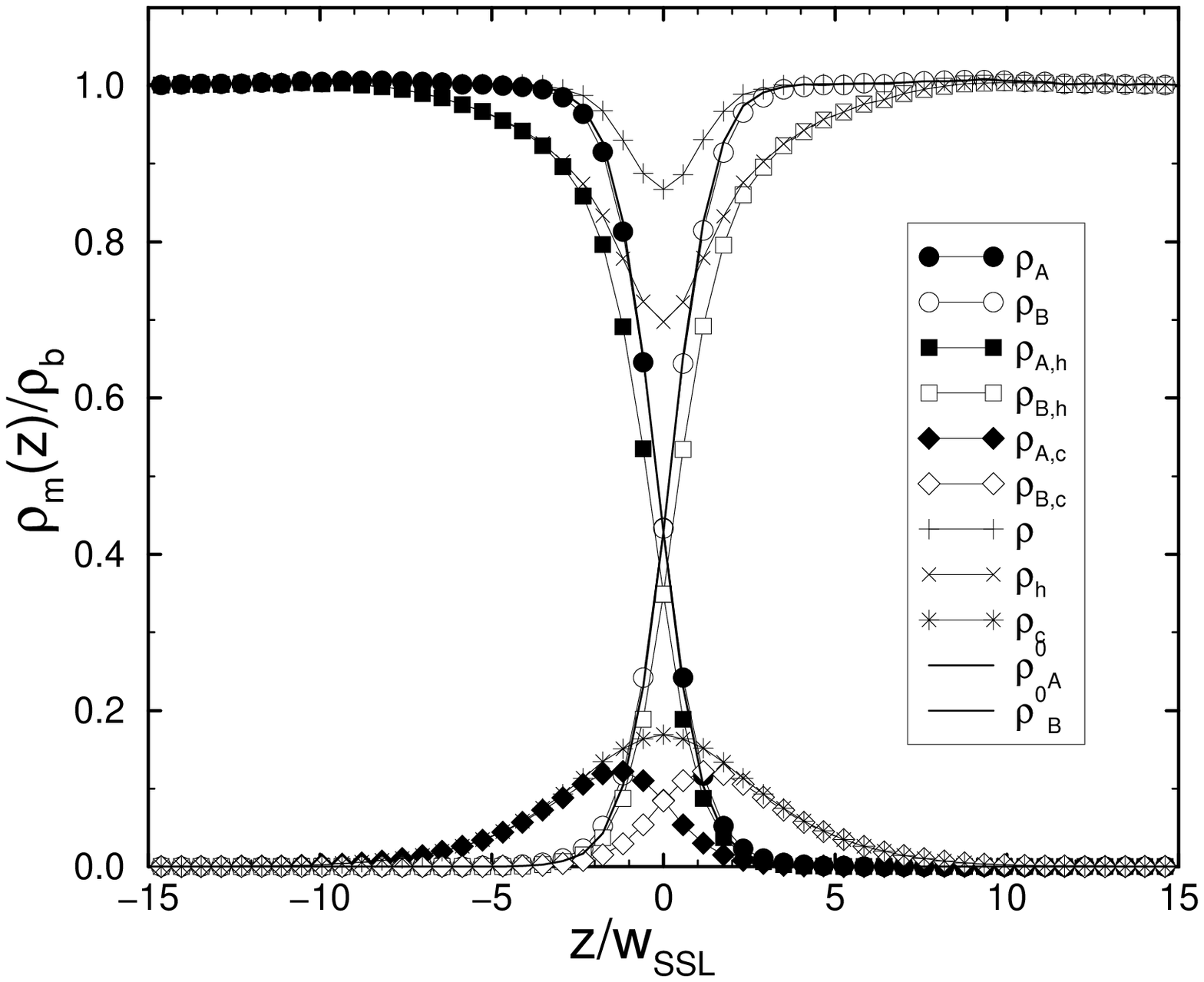}{100}{80}
(b)\fig{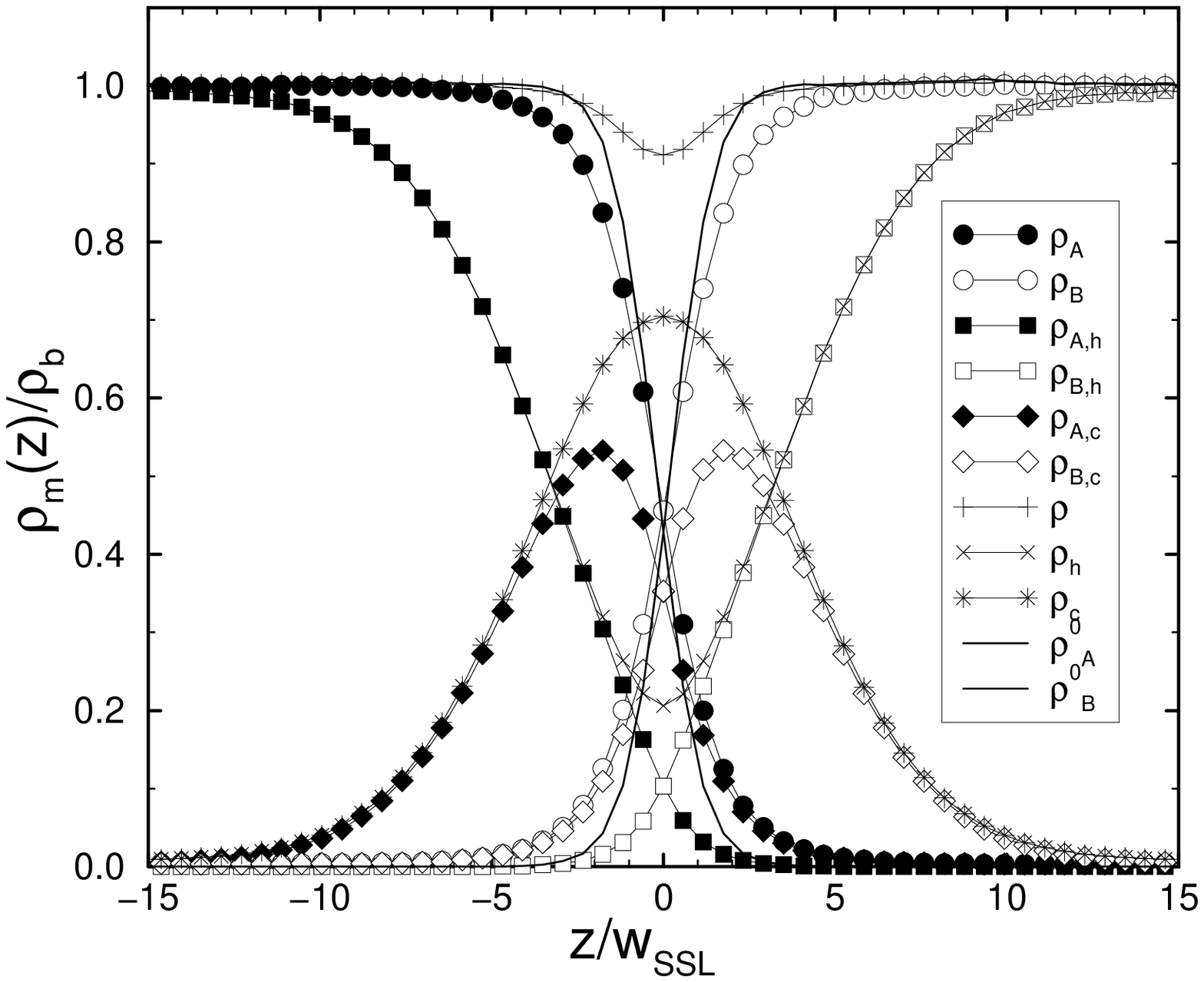}{100}{80}

\begin{figure}
\caption{ \label{fig4}}
Various density profiles at (a) $\delta \mu=0$ and (b) $\delta \mu = 3$
vs. $z$ in units of $w_{\rm SSL}=b/\sqrt{6 \chi}$. Specifically shown are
the total densities of A and B monomers $\rho_A$ and $\rho_B$, the
densities of A and B homopolymer monomers $\rho_{A,h}$ and $\rho_{B,h}$,
the densities of A and B copolymer monomers $\rho_{A,c}$ and $\rho_{B,c}$,
the density of all monomers $\rho$, the density of homopolymer monomers
$\rho_h$, and the density of copolymer monomers $\rho_c$. For comparison,
the density profiles of A and B monomers $\rho_A^0$ and $\rho_B^0$ at
a pure homopolymer interface are also marked. Densities are given in
units of the total bulk density $\rho_b$.
\end{figure}

\clearpage

\noindent
(a)\fig{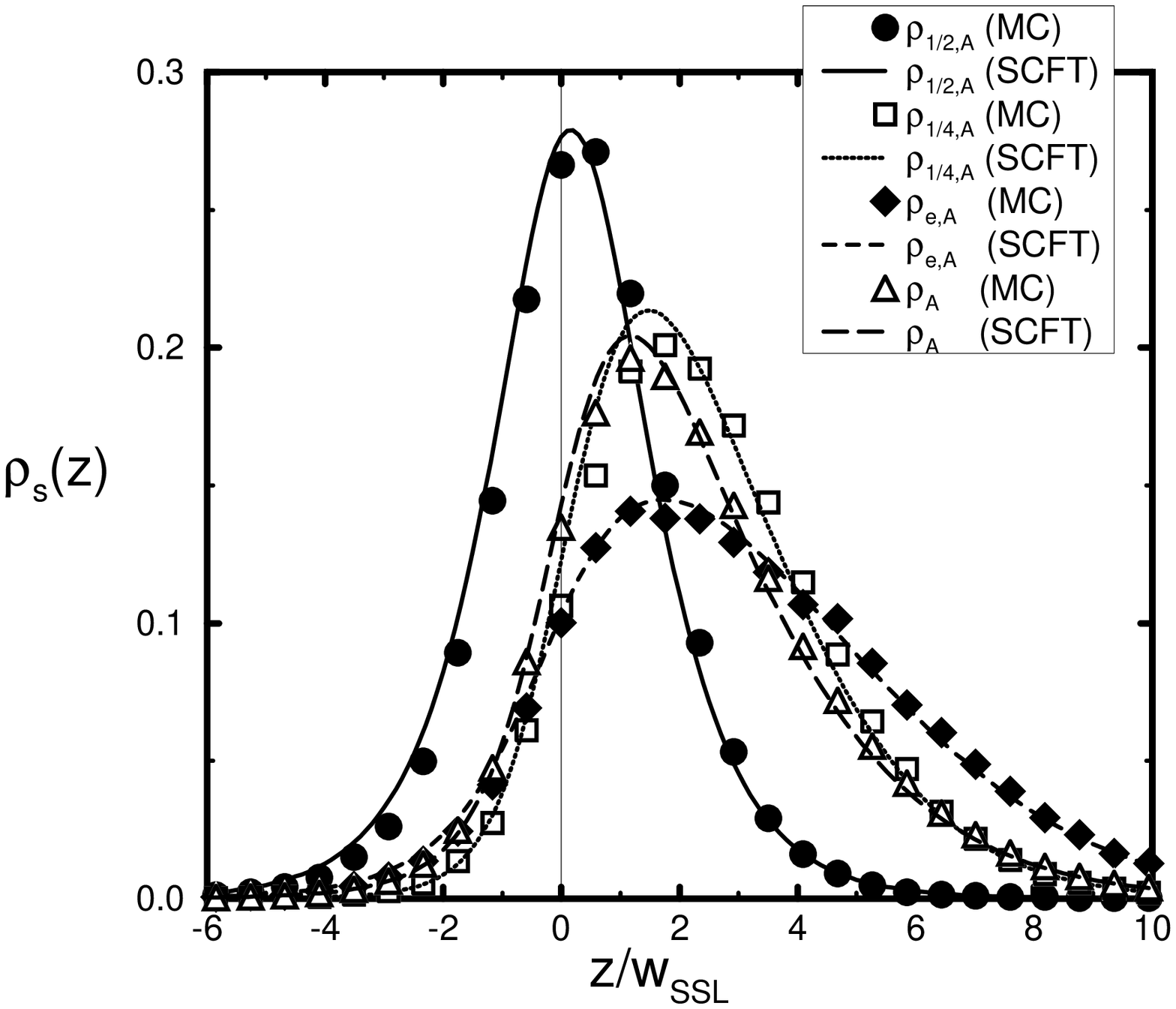}{100}{80}
(b)\fig{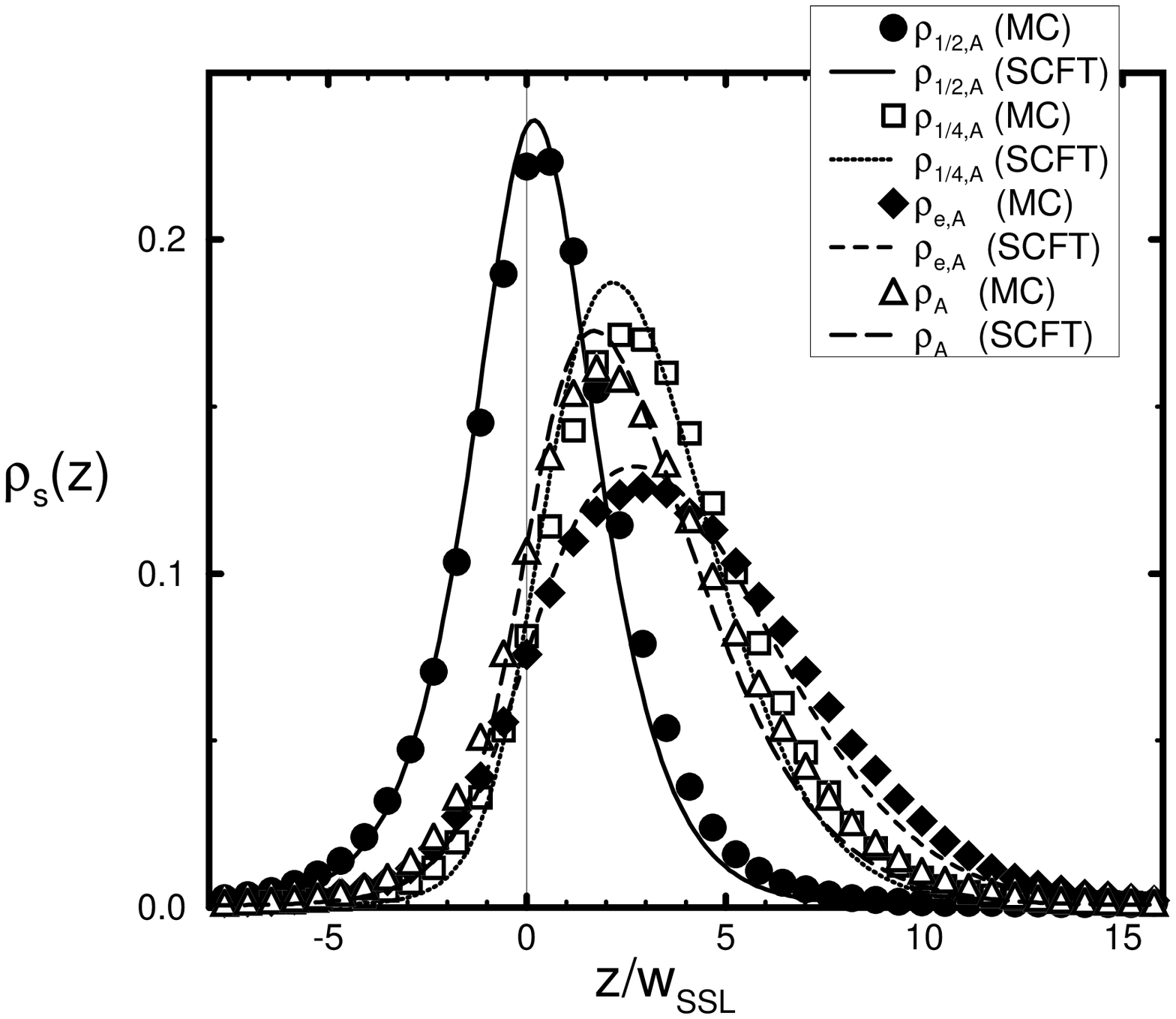}{100}{80}


\begin{figure}
\caption{ \label{fig5}}
Copolymer segment profiles 
vs. $z$ in units of $w_{\rm SSL}=b/\sqrt{6 \chi}$
at (a) $\delta \mu=0$ and (b) $\delta \mu = 3$.
Specifically, profiles are given for the A monomers in the middle of
the chain ($\rho_{1/2,A}$), in the middle of the A block
($\rho_{1/4,A}$), and at the end of the chain ($\rho_e$). For
comparison, the total density of A copolymer monomers is also
shown ($\rho_A$). The lines denote the
self-consistent field results. The profiles are normalized such
that the area under each curve is 1.
\end{figure}


\begin{figure}
\fig{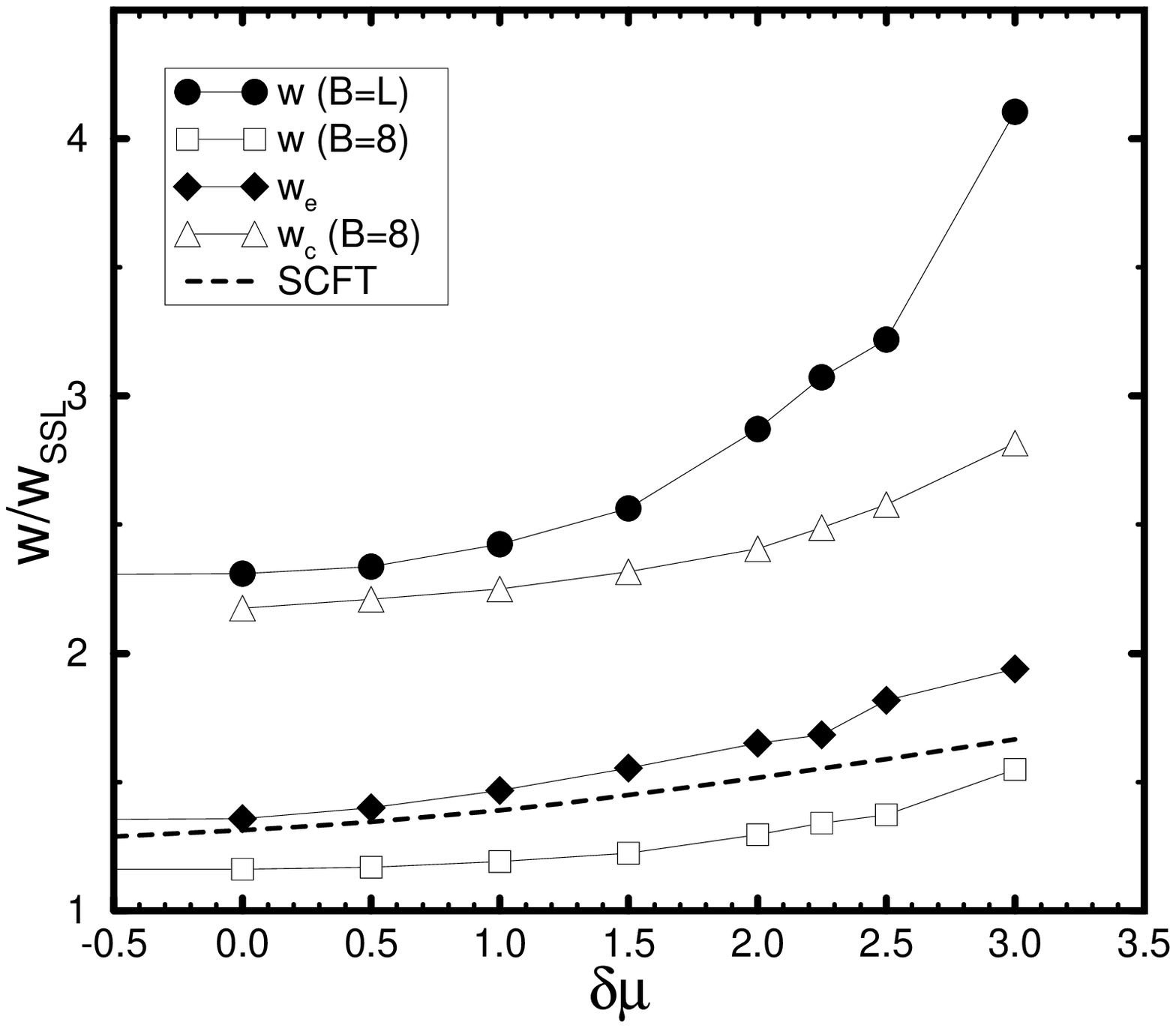}{100}{80}
\caption{ \label{fig6}}
Interfacial width $w$ in units of $w_{\rm SSL} = b/\sqrt{6 \chi}$ as a function
of $\delta \mu$, obtained from different definitions:
From a fit of the profile $m(z)$ to a tanh profile at block size
$B=L=128$ (filled circles) and $B=8$ (open circles), from the width
of the copolymer joint profile (open triangles), and from the excess
internal energy at the interface (filled diamonds, from Ref. [23]). 
The dashed line corresponds to the SCF calculations. See text for
further explanation.
\end{figure}


\begin{figure}
\fig{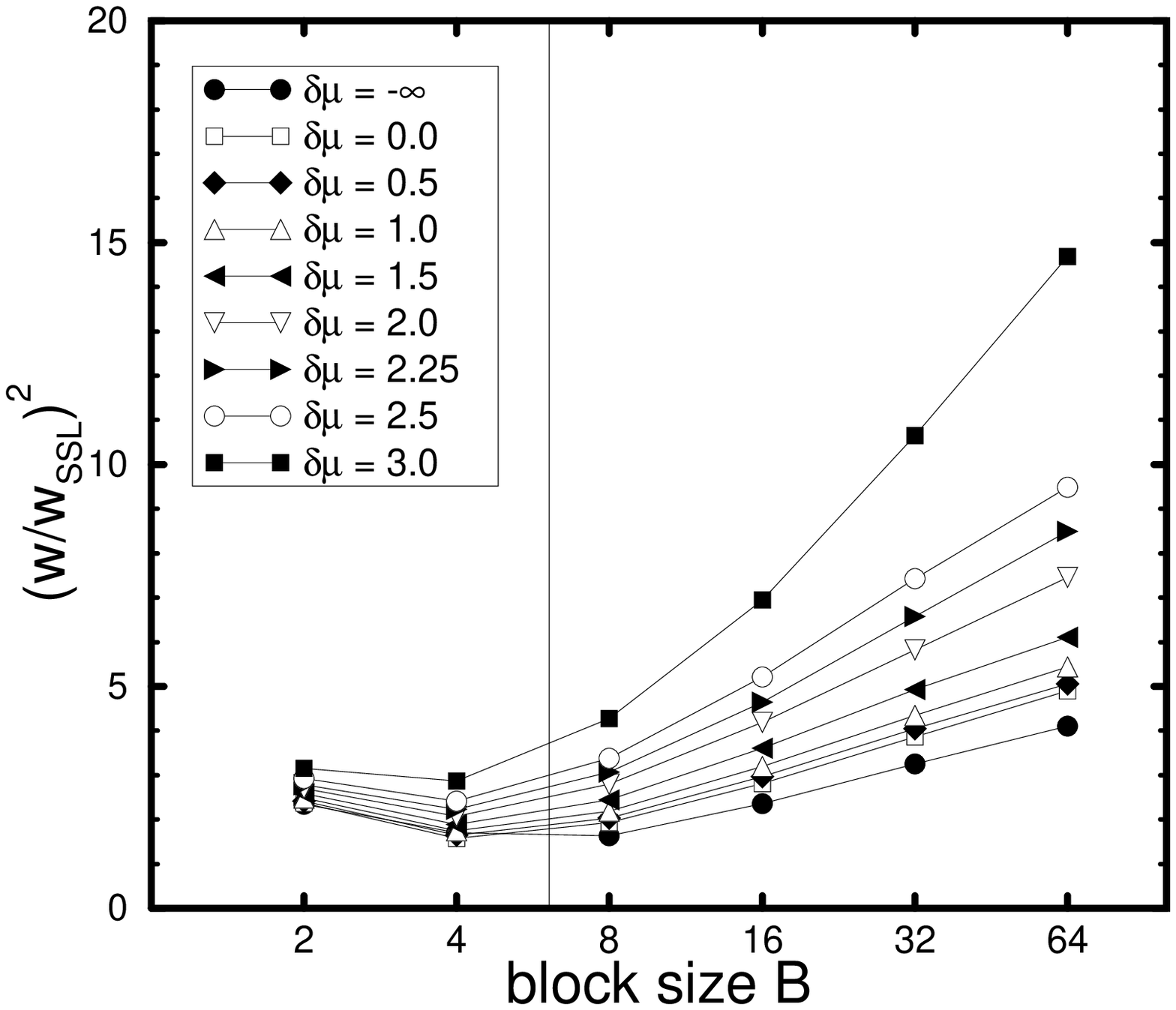}{100}{80}
\caption{ \label{fig7}}
Squared interfacial width $w$ in units of $w_{\rm SSL} = b/\sqrt{6 \chi}$ 
as a function of block size $B$, obtained by fitting $m(z)$ to a
tanh profile, for different values of $\delta \mu$.
\end{figure}


\begin{figure}
\fig{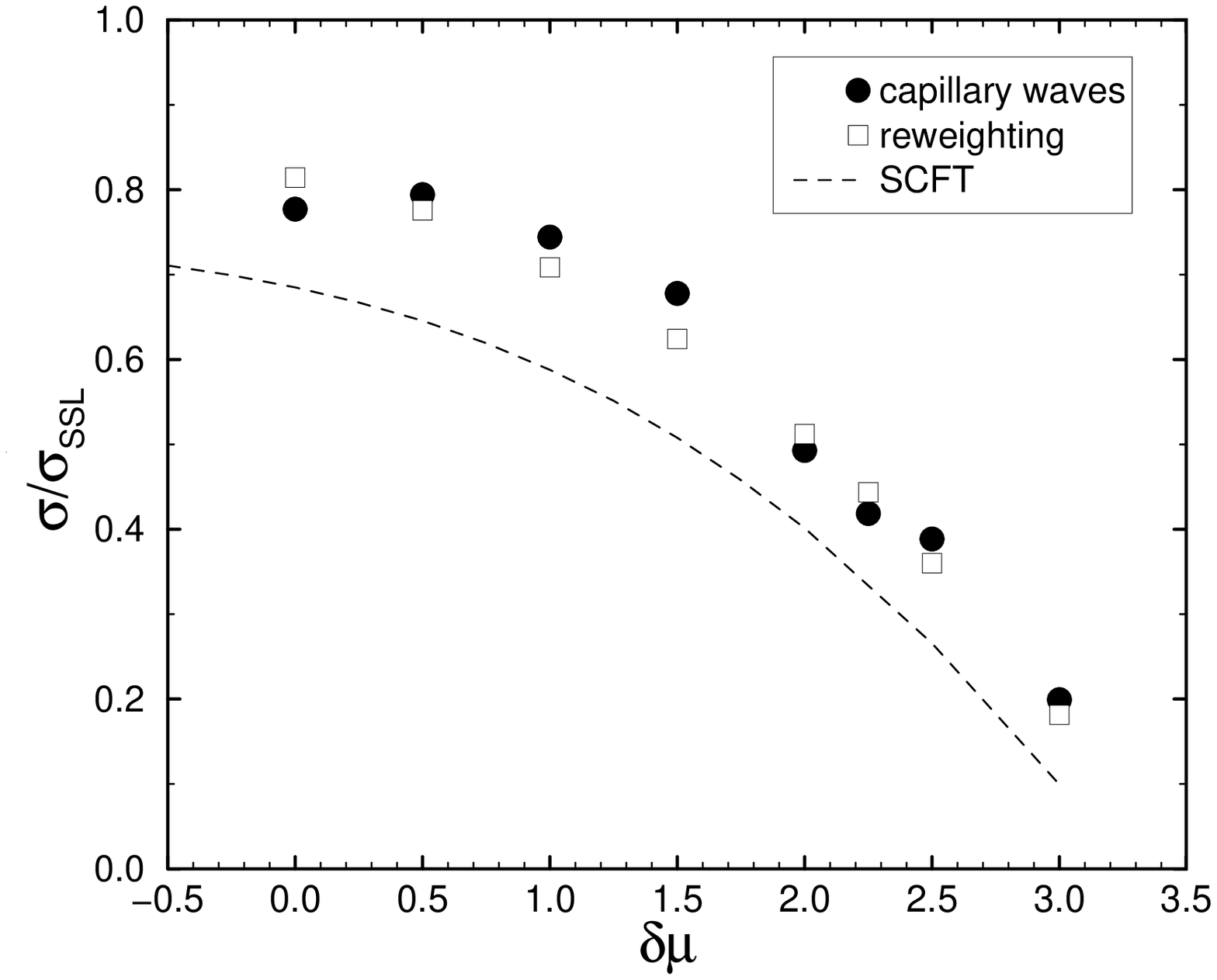}{100}{80}
\caption{ \label{fig8}}
Interfacial tension $\sigma$ in units of 
$\sigma_{\rm SSL} = \sqrt{\chi/6} \; \rho b k_B T$
as a function of $\delta \mu$. Filled circles show data
obtained from the block analysis, open squares show results
from Ref. [23] obtained with a histogram method, and
dashed line marks the prediction of the self-consistent field theory.
\end{figure}


\begin{figure}
\fig{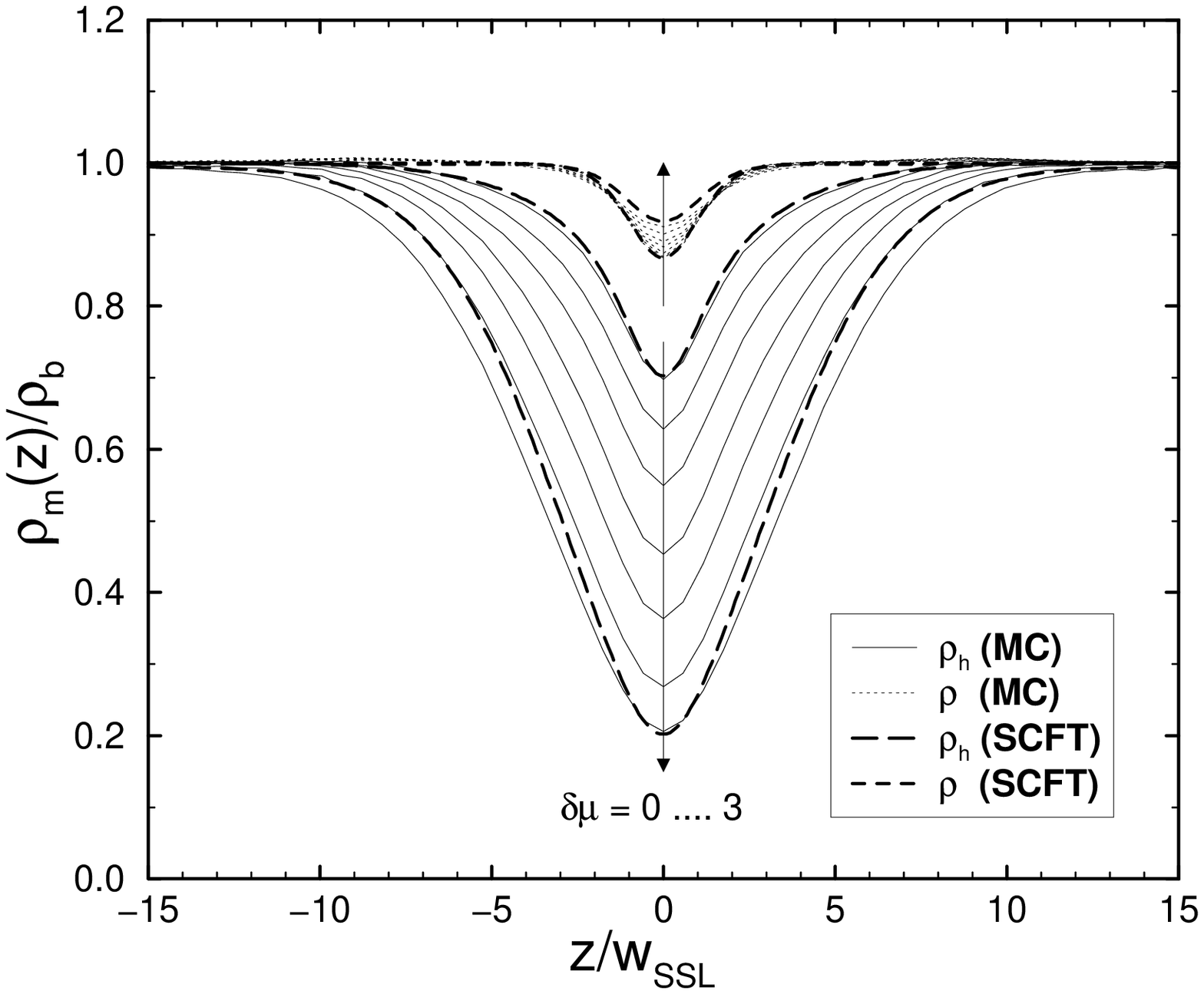}{100}{80}
\caption{ \label{fig9}}
Total monomer density $\rho$ (thin dotted lines)
and homopolymer density $\rho_h$ (thin solid lines) 
in units of $\rho_b$ vs. $z/w_{\rm SSL}$ (with $w_{\rm SSL}=b/\sqrt{6 \chi}$)
for different chemical potentials $\delta \mu$ between 0 and 3
in steps of 0.5. Arrows show the direction of increasing $\delta \mu$.
Thick dashed lines show the prediction of the self-consistent
field theory for $\delta \mu =0$ and $\delta \mu = 3$.
\end{figure}

\clearpage

\noindent
(a)\fig{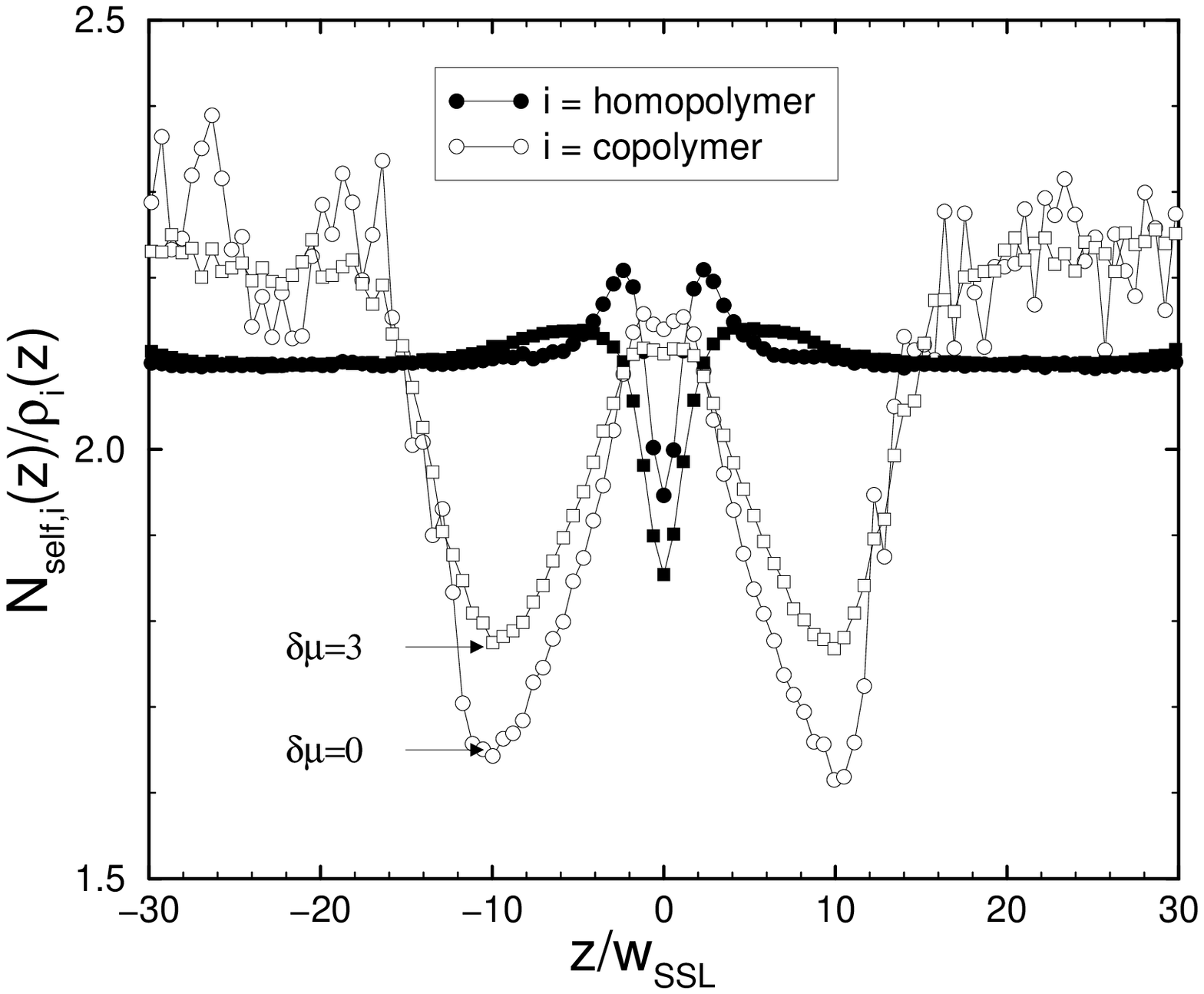}{100}{80}
(b)\fig{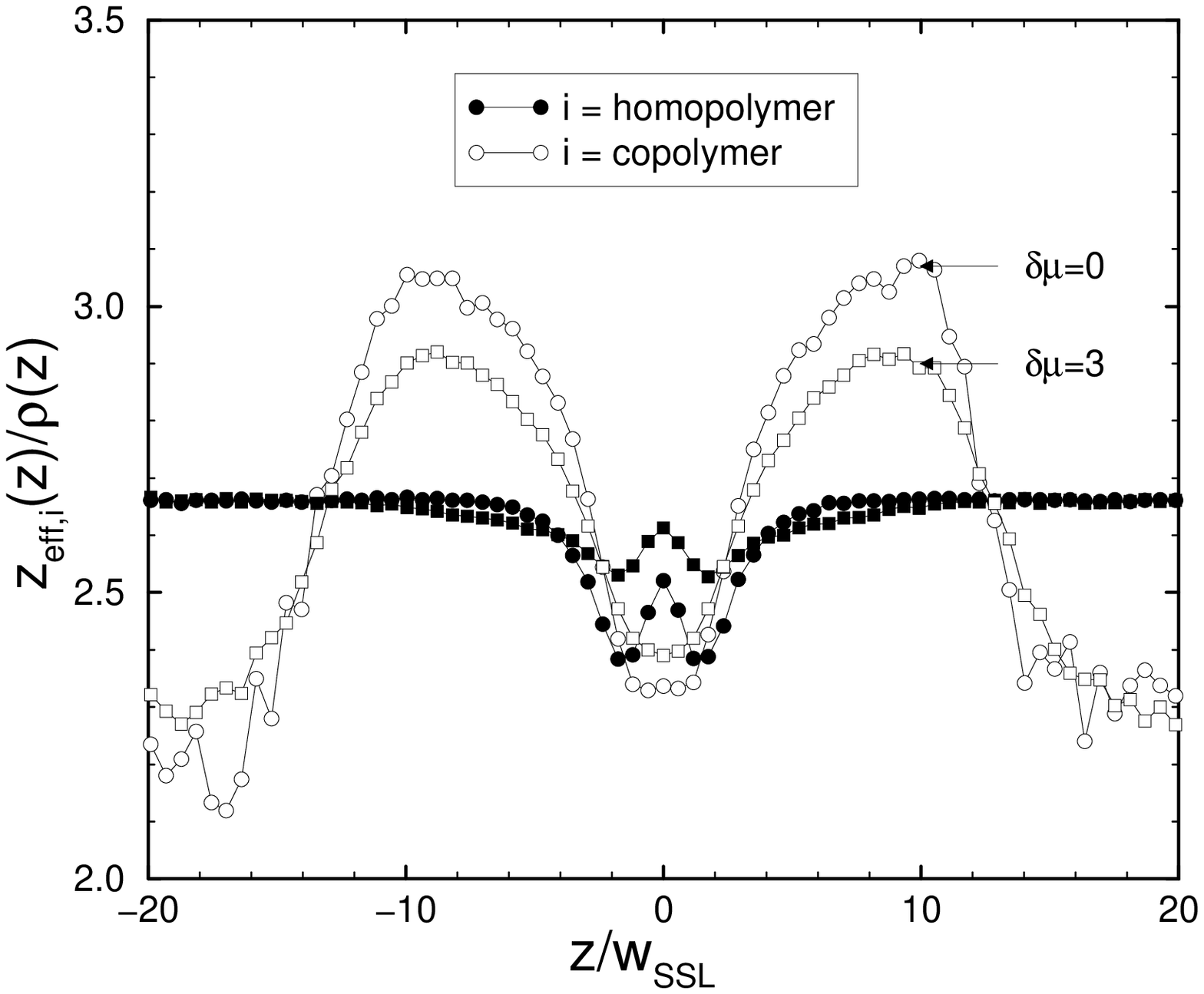}{100}{80}


\begin{figure}
\caption{ \label{fig10}}
Normalized number of self-contacts $N_{\rm self}$ (a) and normalized
effective coordination number $z_{\rm eff}$ (b) vs. $z/w_{\rm SSL}$ with
$w_{\rm SSL}=b/\sqrt{6 \chi}$ for homopolymers and copolymers, and for
$\delta \mu = 0$ and 3. Densities $\rho(z)$ are in units of $\rho_b$.
\end{figure}


\begin{figure}
\fig{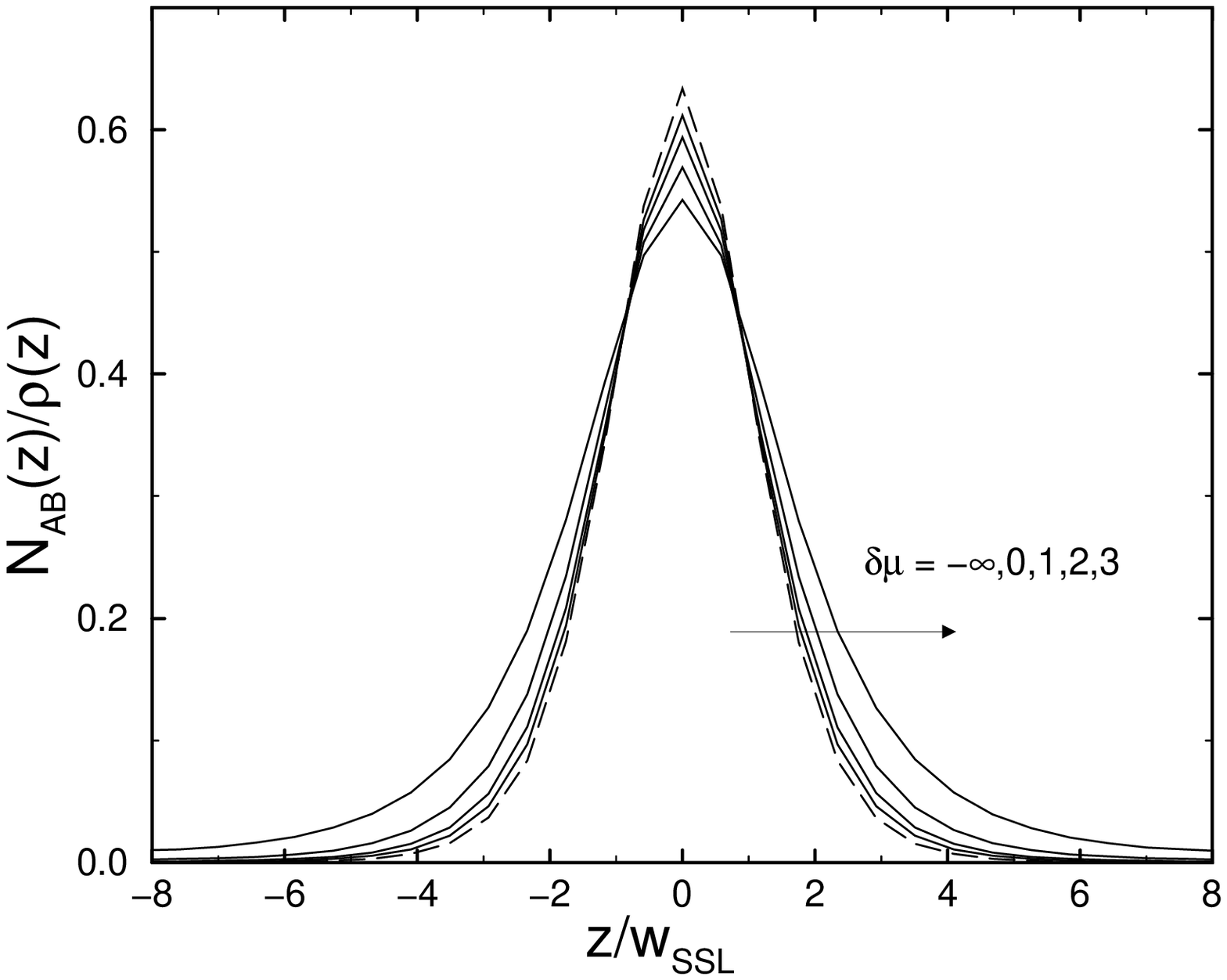}{100}{80}
\caption{ \label{fig11}}
Normalized number of contacts between A and B monomers $N_{AB}$ 
vs. $z/w_{\rm SSL}$ with $w_{\rm SSL}=b/\sqrt{6 \chi}$ for different values of
$\delta \mu$, $\delta \mu \to - \infty$ and $\delta \mu \in [0,3]$ in steps of 
0.5. The arrow indicates
the direction of ascending $\delta \mu$. Inset shows for comparison the 
total number of AB contacts per monomer.
Densities $\rho(z)$ are in units of $\rho_b$.
\end{figure}


\begin{figure}
\fig{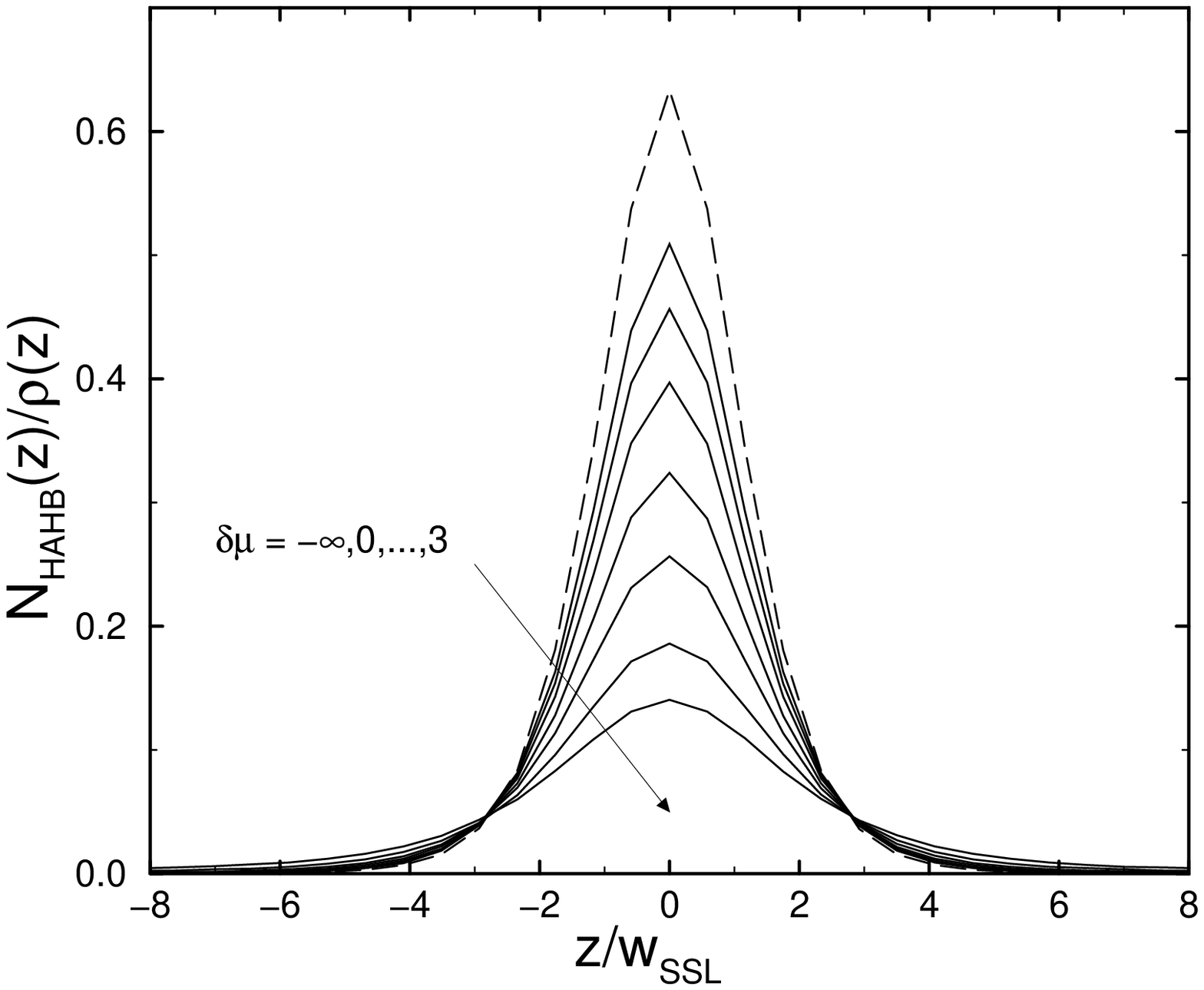}{100}{80}
\caption{ \label{fig12}}
Normalized number of contacts between A and B homopolymer monomers 
$N_{HAHB}$ vs. $z/w_{\rm SSL}$ with $w_{\rm SSL}=b/\sqrt{6 \chi}$ for 
$\delta \mu \to \-\infty$ and $\delta \mu$ between 0 and 3 in steps of 0.5. 
The arrow indicates
the direction of ascending $\delta \mu$. 
Densities $\rho(z)$ are in units of $\rho_b$.
\end{figure}


\begin{figure}
\fig{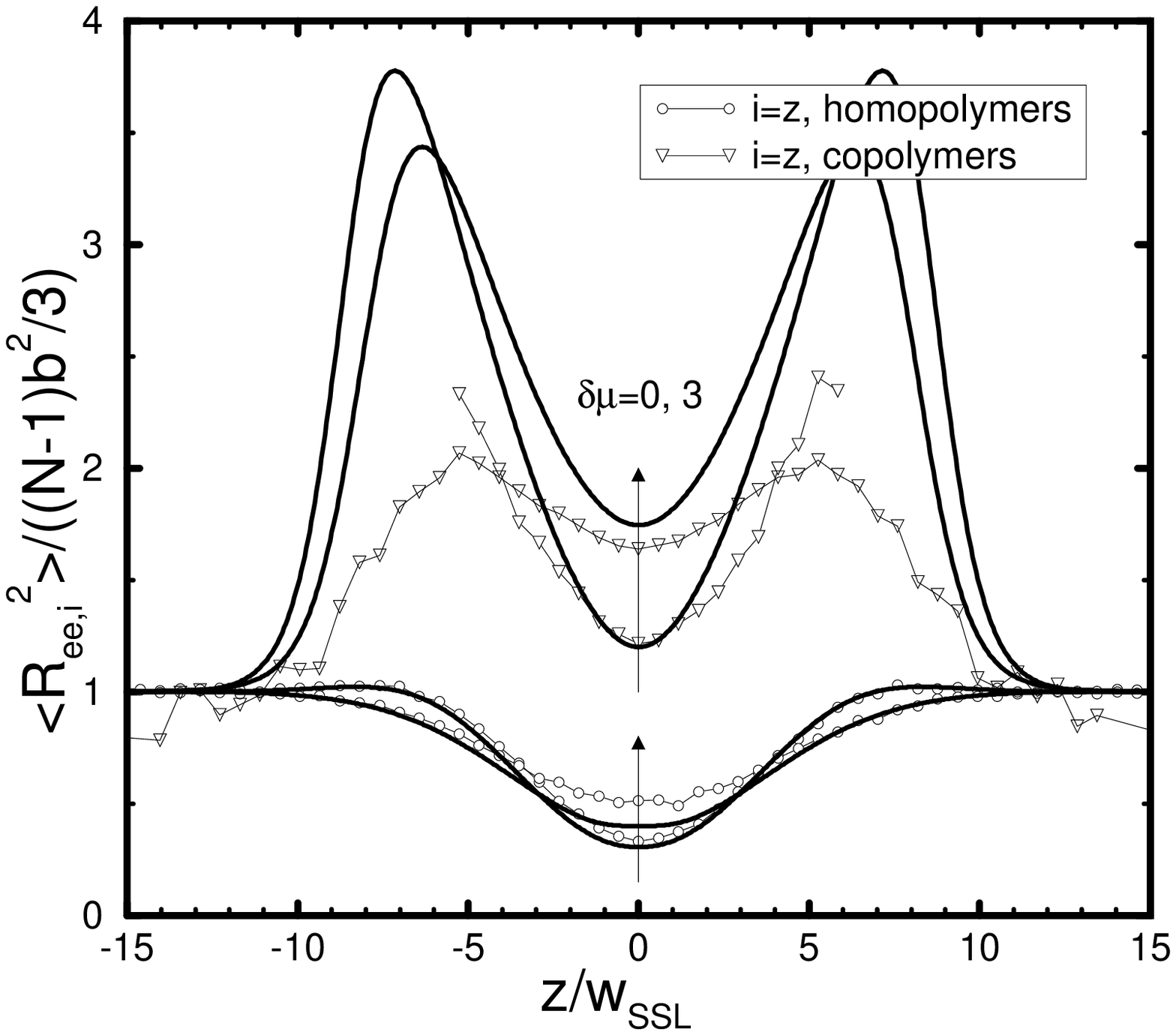}{100}{80}
\caption{ \label{fig13}}
$z$-component of the squared end-to-end vector $R_{ee,z}^2$ in units
of the bulk value, $b^2 (N-1)/3$, vs. the distance $z$ of the
center of the end-to-end vector from the center of the interface
in units of $w_{\rm SSL}=b/\sqrt{6 \chi}$, for copolymers (triangles)
and homopolymers (circles) and $\delta \mu =0,1,2,3$. Lines indicate
the prediction of the self-consistent field theory for 
$\delta \mu=0$ and 3.
\end{figure}


\begin{figure}
\fig{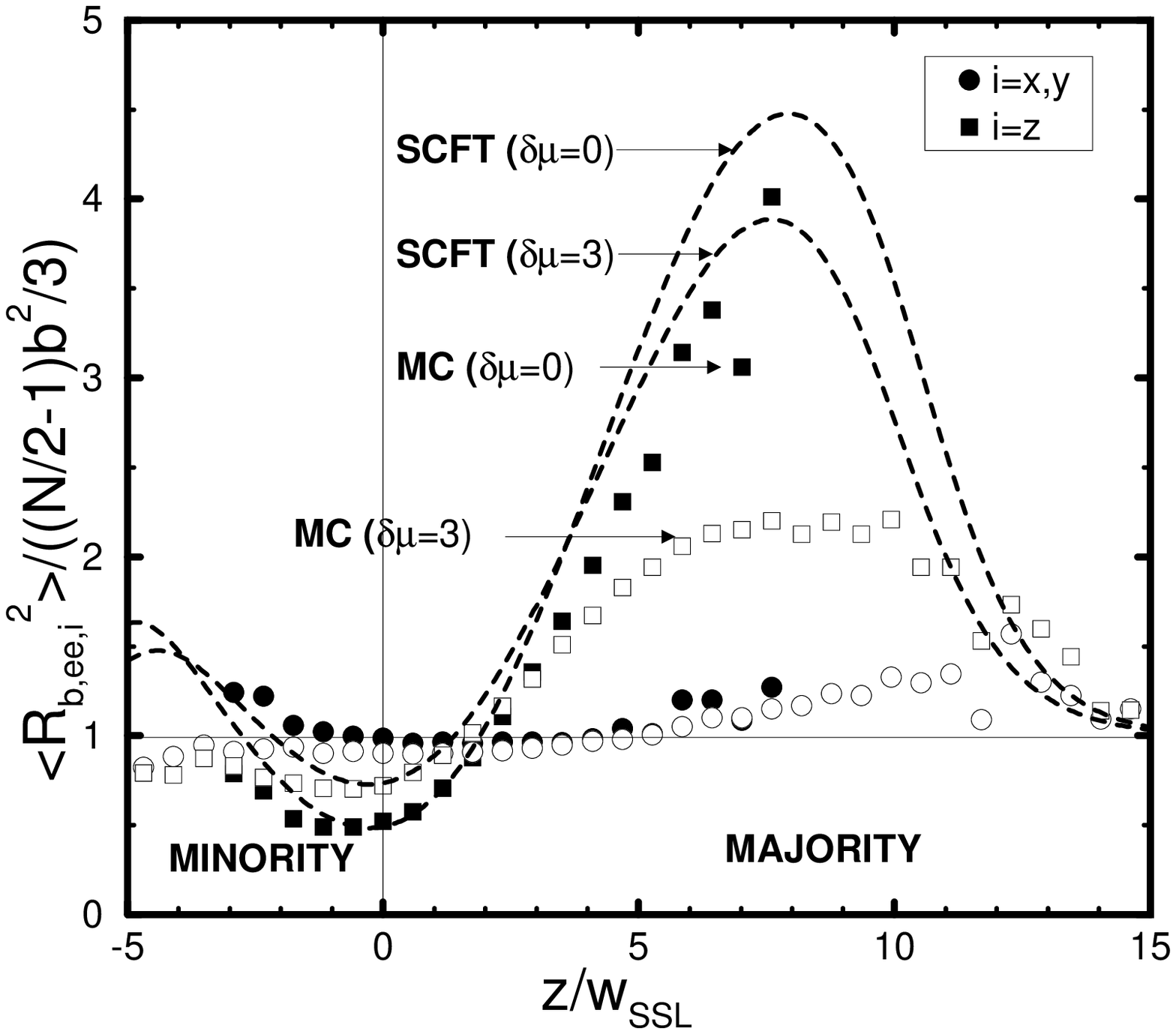}{100}{80}
\caption{ \label{fig14}}
$x$-, $y$-, and $z$-component of the squared end-to-end vector $R_{bee,i}^2$ 
of the A-block in the copolymer, in units of the bulk value 
$b^2 (N/2-1)/3$, vs. the distance $z/w_{\rm SSL}$ of the
center of the end-to-end vector from the center of the interface,
for $\delta \mu =0$ and 3. Units are $w_{\rm SSL}=b/\sqrt{6 \chi}$.
Lines indicate the prediction of the self-consistent field theory.
\end{figure}


\begin{thebibliography}{99}
\bibitem{FN1}
Integrating over this function, one obtains 
$\int_{0}^z \rho_{1/2}(z') dz' \propto  \mbox{Erf}(\sqrt{\Pi} z/(2 w_c)) $,
which resembles a $\tanh (z/w_c)$ profile at the center of the interface
\end{thebibliography}
\end{document}